\documentstyle[12pt]{article}
\begin{document}
\def\t{\times}\def\p{\phi}\def\P{\Phi}\def\a{\alpha}\def\e{\varepsilon}
\def\be{\begin{equation}}\def\ee{\end{equation}}\def\l{\label}
\def\0{\setcounter{equation}{0}}\def\b{\beta}\def\S{\Sigma}\def\C{\cite}
\def\r{\ref}\def\ba{\begin{eqnarray}}\def\ea{\end{eqnarray}}\def\n{\nonumber}
\def\R{\rho}\def\X{\Xi}\def\x{\xi}\def\La{\Lambda}\def\la{\lambda}
\def\d{\delta}\def\s{\sigma}\def\f{\frac}\def\D{\Delta}\def\pa{\partial}
\def\Th{\Theta}\def\o{\omega}\def\O{\Omega}\def\th{\theta}\def\ga{\gamma}
\def\Ga{\Gamma}\def\h{\hat}\def\rar{\rightarrow}\def\vp{\varphi}
\def\inf{\infty}\def\le{\left}\def\ri{\right}\def\foot{\footnote}
\def\un{\underline}\def\ve{\varepsilon}\def\po{\propto}\def\Re{{\rm
Re}}\def\Im{{\rm Im}}

\begin{center}

{\huge\bf Phase transitions signal in inelastic hadron collisions}

\vskip 0.3cm
{\large\it J.Manjavidze,~A.Sissakian}\\
{\large JINR, Dubna}
\end{center}
\vskip 1cm
\begin{abstract}
The consequence of vacuum instability against particles creation
is described using lattice gas model. It is shown that the tail of
multiplicity distribution should decrees slower than any power of
$\exp\{-n\}$, but faster than any power of $1/n$ if the vacuum is
unstable against particles creation.
\end{abstract}

\section{Introduction}\0

The aim of this paper is to find the experimentally observable
consequences of collective phenomena in the high energy hadrons
inelastic collision. We will pay the main attention on the phase
transitions, living out other possible interesting collective
phenomena.

First of all, the statistics experience dictate that we should
prepare the system to the phase transition. The temperature in a
critical domain and the equilibrium media are just this
conditions. It is evident, they are not trivial requirement
considering the hadrons inelastic collision at high energies.

The collective phenomena by definition suppose that the kinetic
energy of particles of media are comparable, or even smaller, than
the potential energy of theirs interaction. It is a quite natural
condition noting that, for instance, the kinetic motion may destroy,
even completely at a given temperature $T$, necessary for the phase
transition long-range order. This gives, more or less definitely, the
critical domain.

The same idea as in statistics seems natural in the multiple
production physics.  We will assume ({\bf A}) that the collective
phenomena should be seen just in the very high multiplicity (VHM)
events, where, because of the energy-momentum conservation laws, the
kinetic energy of created particles can not be high.

We will lean at this point on the $S$-matrix interpretation of
statistics \C{elpat}.  It based on the $S$-matrix generalization of
the Wigner function formalism of Carruzers and Zachariazen \C{carr}
and the real-time finite temperature field theory of Schwinger and
Keldysh \C{sch, kel}.

First of all, the $n$-particle partition function in this approach
coincide with the $n$ particles production cross section $\s_n(s)$
(in the appropriate normalization condition). Secondly, $\s_n(s)$ can
be calculated from the $n$-point Wigner function
$W_n(X_1,X_2,...,X_n)$. In the relativistic case $X_k=(u,q)_k$ are
the 4-vectors. So, the external particles are considered as the
`probes' to measure the interacting fields state, i.e. the low mean
energy of probes means that the system is `cold'.

The multiple production phenomena may be considered also as the
thermalization process of incident particles kinetic energy
dissipation into the created particles mass.  From this point of
view the VHM processes are highly nonequilibrium since the final
state of this case is very far from initial one. It is known in
statistics \C{kac} that such process aspire to be the stationary
Markovian with high level of entropy production. In the case of
complete thermalization the final state is equilibrium.

The equilibrium we will classify as the condition in frame of which
the fluctuations of corresponding parameter are Gaussian. So, in the
case of complete thermalization the probes should have the Gauss
energy spectra.  In other terms, the necessary and sufficient
condition of the equilibrium is smallness of mean value of energy
correlators \C{bog, elpat}. From physical point of view, absence of
this correlators means depression of the macroscopic energy flows in
the system.

The multiple production experiment shows that created particles
energy spectrum is far from Gauss law, i.e. the final states are far
from equilibrium. The natural explanation of this phenomena consist
in presence of (hidden) conservation laws in the interacting
Yang-Mills fields system: it is known that presence of sufficient
number of first integrals in involution prevents thermalization
completely.

But nevertheless the VHM final state may be equilibrium ({\bf B}) in
the above formulated sense. This means that the forces created by the
non-Abelian symmetry conservation laws may be frozen during
thermalization process (remembering its stationary Markovian
character in the VHM domain). We would like to take into account that
the entropy $\cal S$ of a system is proportional to number of created
particles and, therefore, $\cal S$ should tend to its maximum in the
VHM region \C{fermi}.

One may consider following small parameter $(\bar{n}(s)/n)<<1$, where
$\bar{n}(s)$ is the mean value of multiplicity $n$ at given CM energy
$\sqrt s$.  Another small parameter is the energy of fastest hadron
$\e_{max}$. One should assume that in the VHM region $(\e_{max}/\sqrt
s)\to0$.  So, the conditions:
\be
\f{\bar{n}(s)}{n}<<1,~\f{\e_{max}}{\sqrt s}\to0
\l{A}\ee
would be considered as the mark of the processes under consideration.
We can hope to organize the perturbation theory over them
having this small parameters. In this sense VHM processes may be
`simple', i.e. one can use for theirs description semiclassical
methods.

So, considering VHM events one may assume that the conditions ({\bf
A}) and ({\bf B}) are hold and one may expect the phase transition
phenomena. The paper is organized as follows. In Sec.2 we will offer
the qualitative picture of expected phase transition. In sec.3 we
will describe quantitatively this phenomena to find the predictions
for experiment. In Sec.4 we will give formal derivation of
the formalism used in Sec.3.

\section{Condensation and singularity at $z=1$}\0

The $S$-matrix interpretation of statistics is based on following
definitions.  First of all, let us  introduce the generating function
\C{conf}:
\be
T(z,s)=\sum_n z^n \s_n(s).
\l{1}\ee
Summation is performed over all $n$ up to $n_{max}=\sqrt{s}/m$ and,
at finite CM energy $\sqrt{s}$, $T(z,s)$ is a polynomial function of
$z$.  Let us assume now that $z$ is sufficiently small and by this
reason $T(z,s)$ depends on upper boundary $n_{max}$ weakly. In this
case one may formally extend summation up to infinity and in this
case $T(z,s)$ may be considered as a whole function.  This
possibility is important being equivalent of thermodynamical limit
and it allows to classify the asymptotics over $n$ in accordance with
position of singularities over $z$.

Let us consider $T(z)$ (prove will be given below) as the big
partition function, where $z$ is `activity'. It is known \C{lee} that
$T(z)$ should be regular in the circle of unite radii. The leftist
singularity lie at $z=1$. This singularity is manifestation of the
first order phase transition \C{lee, kac, langer}.

The origin of this singularity was investigated carefully in the
paper \C{kac}. It was shown that position of singularities over $z$
depends on the number of particles $n$ in the system: the two complex
conjugated singularity moves to the real $z$ axis with rising $n$
and in the thermodynamical limit $n=\infty$ they pinch point $z=1$.
More careful analysis \C{langer} shows that if the system is
equilibrium then $T(z)$ may be singular at $z=1$ and $z=\infty$ only.

The position of singularity over $z$ and the asymptotic behavior
of $\s_n$ are related closely. Indeed, for instance, inserting into
(\r{1}) $\s_n\propto \exp\{-cn^\ga\}$ we find that $T(z)$ is singular
at $z=1$ if $\ga<1$. Generally, using Mellin transformation,
\be
\s_n=\f{1}{2\pi i}\oint\f{dz}{z^{n+1}}T(z)
\l{1.1}\ee
This integral can be calculated expanding it in vicinity of $z_c$,
where $z_c$ is smallest real positive solution of equation:
\be
n=z\f{\pa}{\pa z}\ln T(z).
\l{1.2}\ee
Then integral (\r{1.1}) have following estimation:
\be
\s_n\propto e^{-n\ln z_c(n)},~z_c>1.
\l{1.3}\ee
Therefore, to have the singularity at $z=1$ we should consider
$z_c(n)$ as a decreasing function of $n$. On other hand, at
constant temperature, $\ln z_c(n)\sim\mu_c(n)$ is the chemical
potential, i.e. is a work necessary for one particle creation. So,
the singularity at $z=1$ means that the system is unstable: the less
work is necessary for creation of one more particle if $\mu(n)$ is
the decreasing function of $n$.

The physical explanation of this phenomena is following, see also
\C{andr}. Generating function $T(z)$ have following expansion:
\be
T(z)=\exp\{\sum_lz^lb_l\},
\l{1.4}\ee
where $b_l$ are known as the Mayer's group coefficients. They
can be expressed through the inclusive correlation functions and may
be used to describe formation of droplets of correlated particles.
So, if droplet consist from $l$ particles, then
\be
b_l\sim e^{-\b\xi l^{(d-1)/d}}
\l{1.5}\ee
is the mean number of such droplets. Here $\xi l^{(d-1)/d}$ is the
surface energy of $d$-dimensional droplet.

Inserting this estimation into (\r{1.4}),
\be
\ln T(z)\sim\sum_l e^{\b(l\mu -\xi l^{(d-1)/d})},~\b\mu=\ln z.
\l{1.6}\ee
First term in the exponent is the volume energy of droplet and being
positive it try to enlarge the droplet. The second surface term try
to shrink it. Therefore, singularity at $z=1$ is the consequence of
instability: at $z>1$ the volume energy abundance leads to unlimited
grow of the droplet.

\section{Condensation and type of asymptotics over multiplicity}\0

It is important for VHM experiment to have upper restriction on the
asymptotics. We wish to show that $\s_n$ decrease faster than any
power of $1/n$:
\be
\s_n<O(1/n).
\l{2.1}\ee
To prove this estimation one should know the type of singularity at
$z=1$. The detailed derivation of the model used for this purpose we
will give in subsequent section.

One can imagine that the points, where the external particles are
created, form the gas system. Here we assume that this system is
equilibrium, i.e. there is not in this system macroscopical flows
of energy, particles, charges and so on.

The lattice gas approximation is used to describe such system. This
description is quite general and did not depend on details. Motion of
the gas particles leads to necessity sum over all distributions of
the particles on cells. For simplicity we will assume that only
one particle can occupy the cell.

So, we will introduce the occupation number $\s_i=\pm1$ in the $i$-th
cell: $\s_i=+1$ we have not particle in the cell and $\s_i=-1$ means
that the particle exist. Assuming that the system is equilibrium we
may use the ergodic hypothesis and sum over all `spin' configurations
of $\s_i$, with restriction: $\s_i^2=1$. It is well known \C{wilson}
this restriction introduce the interactions.

Corresponding partition function in temperature representation
\C{langer}
\be
\R(\b,H)=\int D\s e^{-S\la(\s)}
\l{2.2}\ee
where integration is performed over $|\s(x)|\leq\infty$ and,
considering the continuum limit, $D\s=\prod_x d\s(x)$. The action
\be
S_\la(\s)=\int dx\le\{\f{1}{2}(\nabla \s)^2 -\o\s^2+g\s^4-\la\s\ri\}
\l{2.3}\ee
where
\be
\o\sim\le(1-\f{\b_{cr}}{\b}\ri),~g\sim\f{\b_{cr}}{\b},~
\la\sim\le(\f{\b_{cr}}{\b}\ri)^{1/2}\b H.
\l{2.4}\ee
and $1/\b_{cr}$ is the critical temperature.

\subsection{Unstable vacuum}

We start consideration from the case $\o>0$, i.e. assuming that
$\b>\b_{cr}$.  In this case the ground state is degenerate if $H=0$.
The extra term $\sim\s H$ in (\r{2.3}) can be interpreted as the
interaction with external magnetic field $H$. This term regulate
number of `down' spins with $\s=-1$ and is related to the
activity:
\be
z^{1/2}=e^{\b H},
\l{2.5}\ee i.e. $H$ coincide with chemical potential.

The potential
\be
v(\s)=-\o\s^2+g\s^4,~\o>0,
\l{2.6}\ee
has two minimums at
$$
\s_\pm=\pm\sqrt{\o/2g}.
$$
If the dimension $d>1$ no tunnelling phenomena exist. But choosing
$H<0$ the system in the right minimum (it correspond to the state
without particles) becomes unstable. The system tunneling into the
state with absolute minimum of energy.

The partition function $\R(\b,z)$ becomes singular at $H=0$ because
of this instability. The square root branch point gives
\be
{\rm Im}\R(b,z)=\f{a_1(\b)}{H^4}e^{-a_2(\b)/H^2},~a_i>0.
\l{2.7}\ee
Note, ${\rm Im}\R(b,z)=0$ at $H=0$. Deforming contour of
integration in (\r{1.1}) on the branch line,
\be
\R_n(\b)=\f{1}{\pi}\int^\infty_1\f{dz}{z^{n+1}}\f{8a_1\b^4}{\ln^4z}
e^{-4a_2\b^2/\ln^2z}.
\l{2.8}\ee
In this integral
\be
z_c\propto \exp\le\{\f{8a_2\b^2}{n}\ri\}^{1/3}
\l{2.9}\ee
is essential. This leads to following estimation:
\be
\R_n\propto e^{-3(a_2\b^2)^{1/3}1/3n^{2/3}}<O(1/n).
\l{2.10}\ee

It is useful to note at the end of this section that\\
(i) The value of $\R_n$ is defined by ${\rm Im}\R(b,z)$ and the
metastable states, decay of which gives contribution into ${\rm
Re}\R(b,z)$, are not important.\\
(ii) It follows from (\r{2.9}) that in the VHM domain
\be
H\sim H_c\sim\ln z_c\sim (1/n)^{1/3}\to0.
\l{2.11}\ee
So, the calculations are performed for the week external field case,
when the degeneracy is weekly broken. It is evident that the life
time of the unstable (without particles) state is large in this case
and by this reason the used semiclassical approximation is rightful.
This is important consequence of (\r{A}).

\subsection{Stable vacuum}

Let us consider now $\o<0$, i.e. $\b\b_{cr}$. Potential (\r{2.6})
have only one minimum at $\s=0$ in this case. Inclusion of external
field shifts the minimum to the point $\s_c=\s_c(H)$. In this
case the expansion in vicinity of $\s_c$ should be useful. In result,
\be
\R(\b,z)=\exp\{\int dx \la\s_c-W(\s_c)\},
\l{2.12}\ee
where $W(\s_c)$ can be expanded over $\s_c$:
\be
W(\s_c)=\sum_l\f{1}{l}\int\prod_k\{dx_k\s_c(x_k;H)\}\tilde{b}_l
(x_1,...,x_l).
\l{2.13}\ee
In this expression $\tilde{b}_l(x_1,...,x_l)$ is the one-particle
irreducible $l$-point Green function, i.e. $\tilde{b}_l$ is the
virial coefficient. Then $\s_c$ can be considered as the effective
activity of the correlated $l$-particle group.

The sum in (\r{2.13}) should be convergent and, therefore,
$|s_c|\to\infty$ if $|H|\to\infty$. But in this case the virial
decomposition is equivalent of the expansion over inverse density of
particles \C{virial}. In the VHM region it is high and the mean field
approximation becomes rightful. In result,
\be
\s_c\simeq-\le(\f{|\la|}{4g}\ri)^{1/3}:~|s_c|\to\infty~{\rm if}~
|\la|\to\infty,
\l{2.15}\ee
and
\be
\R(\b,z)\propto e^{\f{3|\la|^{4/3}}{(4g)^{1/3}}}
\le\{12g\le(\f{|\la|}{4g}\ri)^{2/3}\ri\}^{-1/2}.
\l{2.16}\ee
We can use this expression to calculate $\R_n$. In this case
\be
z_c\propto e^{4gn^3}\to\infty~{\rm at}~n\to\infty,
\l{2.17}\ee
is essential and in the VHM domain
\be
\R_n\propto e^{-gn^4}<O(e^{-n}).
\l{2.18}\ee
This result is evident consequence of vacuum stability.
It should be noted once more that the conditions (\r{A}) considerably
simplify calculations.

\section{Conclusion}\0

In conclusion we wish to formulate once more the main assumptions.

(I). It was assumed first of all that the system under consideration
is equilibrium. This condition may be naturally reached in the
statistics, where one can wait the arbitrary time till the system
becomes equilibrium. Note, in the critical domain the time of
relaxation $t_r\sim (T_c/(T-T_c))^\nu$, $(T-T_c)\to+0$, $\nu>0$, $T_c$
is the critical temperature.

We can not give the guarantee that in the high energy hadron
collisions the final state system is equilibrium. The reason of this
uncertainty is the finite time the inelastic processes and presence
of hidden (confinement) constraints on the dynamics.

But we may formulate the quantitative conditions, when the
equilibrium is hold \C{bog}. One should have the Gauss energy spectra
of created particles. If this condition is hardly investigated on the
experiment, then one should consider the relaxation of `long-range'
correlations. This excludes usage of relaxation condition for the
`short-range' (i.e. resonance) correlation

(II). The second condition consist in requirement that the system
should be in the critical domain, where the (equilibrium)
fluctuations of system becomes high.  Having no theory of hadron
interaction at high energies we can not define where is lie the
`critical domain' and even exist it or not.

But anyway, having the VHM `cold' final state we can hope that the
critical domain is achieved. Moreover, noting that the entropy reach
its maximum at given incident energy, we can hope that the VHM state
is equilibrium.

{\bf Acknowledgments}
\vskip 0.5cm

We are grateful to V.G.Kadyshevski for interest to discussed in the
paper questions. We would like to note with gratitude that the
discussions with E.Kuraev was interesting and important.

\end{document}